\documentclass[aps,prb,showpacs,twocolumn,superscriptaddress,floatfix,10pt,longbibliography]{revtex4-1}

\usepackage{graphicx}
\usepackage{amssymb,amsmath} 
\usepackage{color}
\usepackage[normalem]{ulem}
\usepackage{pifont}
\usepackage[colorlinks, urlcolor=blue]{hyperref}
\usepackage[load=abbr]{siunitx}
\usepackage[]{units}

\RequirePackage[xcolornames,svgnames,dvipsnames,rgb]{xcolor}
\pdfpageattr {/Group << /S /Transparency /I true /CS /DeviceRGB>>}

\newcommand{\sic}{{SiC($0001$)}}

\usepackage{bm}
\renewcommand{\vec}[1]{\bm{#1}}

\newcommand*{\Scale}[2][4]{\scalebox{#1}{$#2$}}%

\setcitestyle{numbers,square}

\begin{document}

\title{Structural and electronic properties of Li intercalated graphene on SiC(0001)}
\author{Nuala M. Caffrey}
\affiliation{Department of Physics, Chemistry and Biology (IFM), Link\"oping 
University, SE-581 83, Link\"oping, Sweden}
\email[Electronic address: ]{nuala.mai.caffrey@liu.se}
\author{Leif I. Johansson}
\affiliation{Department of Physics, Chemistry and Biology (IFM), Link\"oping 
University, SE-581 83, Link\"oping, Sweden}
\author{Chao Xia}
\affiliation{Department of Physics, Chemistry and Biology (IFM), Link\"oping 
University, SE-581 83, Link\"oping, Sweden}
\author{Rickard Armiento}
\affiliation{Department of Physics, Chemistry and Biology (IFM), Link\"oping 
University, SE-581 83, Link\"oping, Sweden}
\author{Igor A. Abrikosov}
\affiliation{Department of Physics, Chemistry and Biology (IFM), Link\"oping 
University, SE-581 83, Link\"oping, Sweden}
\affiliation{Materials Modeling and Development Laboratory, NUST ``MISIS'', 
119049 Moscow, Russia}
\affiliation{LACOMAS Laboratory, Tomsk State University, 634050, Tomsk, Russia}
\author{Chariya Jacobi}
\affiliation{Department of Physics, Chemistry and Biology (IFM), Link\"oping 
University, SE-581 83, Link\"oping, Sweden}

\date{\today}

\begin{abstract}
We investigate the structural and electronic properties of Li-intercalated monolayer graphene on SiC(0001) using combined angle-resolved photoemission spectroscopy and first-principles density functional theory. Li intercalates at room temperature both at the interface between the buffer layer and SiC and between the two carbon layers. The graphene is strongly $n$-doped due to charge transfer from the Li atoms and two $\pi$-bands are visible at the $\bar{K}$-point. After heating the sample to 300$^\circ$C, these $\pi$-bands become 
sharp and have a distinctly different dispersion to that of Bernal-stacked bilayer graphene. We suggest that the Li atoms intercalate between the two carbon layers with an ordered structure, similar to that of bulk LiC$_6$. An AA-stacking of these two layers becomes energetically favourable. The $\pi$-bands around the $\bar{K}$-point closely resemble the calculated band structure of a C$_6$LiC$_6$ system, where the intercalated Li atoms impose a super-potential on the graphene electronic structure that opens pseudo-gaps at the Dirac points of the two $\pi$-cones.
\end{abstract}


\maketitle

\section{Introduction}

The epitaxial growth of graphene on SiC substrates is a viable and attractive way of producing large homogeneous monolayer samples suitable for electronic device applications \cite{PhysRevB.78.245403, 
emtsev2009towards}. When grown on the Si terminated SiC(0001) 
surface, the first carbon buffer layer has a detrimental effect on the graphene 
charge carrier mobility and needs to be eliminated \cite{pallecchi2014high}. 
Intercalation of elements such as hydrogen \cite{PhysRevLett.103.246804,  
virojanadara2010buffer}, gold \cite{PhysRevB.81.235408}, germanium 
\cite{emtsev2011ambipolar}, silicon \cite{PhysRevB.85.045418, 
oida2014controlled}, nitrogen \cite{PhysRevB.92.081409}, fluorine 
\cite{walter2011highly}, oxygen \cite{oliveira2013formation}, lithium 
\cite{virojanadara2010epitaxial, virojanadara2010low, watcharinyanon2012studies} 
and sodium \cite{watcharinyanon2012changes, PhysRevB.85.125410, xia2013detailed} 
have all been shown to decouple the buffer layer and transform it into a quasi 
free-standing graphene layer with varying degrees of doping depending on the intercalant. 
The alkali metals induce a strong $n$-type doping when deposited on graphene 
samples. Potassium \cite{bostwick2010observation, PhysRevB.84.085410}, 
rubidium and cesium \cite{watcharinyanon2011rb} remain on the surface after deposition and do not 
penetrate to the interface even after heating. Lithium 
\cite{virojanadara2010epitaxial, virojanadara2010low, watcharinyanon2012studies} 
and sodium \cite{watcharinyanon2012changes, PhysRevB.85.125410, xia2013detailed}, on the other hand, 
intercalate readily at room temperature, decoupling the buffer layer. Increasing the temperature to 300$^\circ$C (100$^\circ$C) promotes the complete intercalation of Li (Na) while higher temperatures result in the de-intercalation and desorption of the metal from the sample. 
Interest in Li-intercalated graphene was originally motivated by the possibility of using it to improve the capacity of Li-ion batteries. However, the structural degradation of the carbon layers after Li deposition renders this unlikely \cite{virojanadara2010low}. More recently, superconductivity was discovered to occur in Li decorated monolayer graphene on SiC(0001) at low temperatures \cite{profeta2012phonon, ludbrook2015evidence}. Detailed studies of the structural and electronic properties of Li intercalated graphene will be required in order to determine the origin and mechanism of this behavior.

Low-energy electron microscopy (LEEM) imaging provide evidence that Li atoms form small islands on the graphene surface directly after deposition \cite{virojanadara2010low}. Over time these islands coalesce, shrink and eventually disappear; this process can be accelerated by annealing. Core level spectra, combined with angle-resolved photoemission spectroscopy (ARPES) data, show that Li intercalates both at the interface between 
the substrate and the buffer layer and between the carbon layers, with the 
result that multiple $\pi$-bands become visible \cite{virojanadara2010epitaxial, 
watcharinyanon2012studies}. A $(\sqrt{3}\times\sqrt{3})R30^\circ$ diffraction pattern was observed after Li deposition on monolayer graphene using low-energy electron diffraction measurements. This pattern did not appear when when Li was deposited on a sample containing only the ($6\sqrt{3}\times6\sqrt{3})R30^\circ$ carbon buffer layer, suggesting that it is due to an ordered intercalated Li layer between the two graphene sheets.
The decoupling of the buffer layer was also shown to occur when Li was deposited 
on samples with an initial coverage of only the buffer layer. In this case, a single $\pi$-band becomes visible directly after Li deposition \cite{virojanadara2010low}, which can only occur if the buffer layer 
has been decoupled.

Several recent investigations have attempted to elucidate the exact nature of 
the Li intercalation on both zero- and mono-layer graphene and its temperature 
dependence. Bisti et al.~determined that the Li atoms occupy the T4 site of the topmost SiC bilayer after they intercalate underneath the buffer layer, thereby decoupling it from the substrate \cite{PhysRevB.91.245411}. 
Sugawara et al.~deposited Li on bilayer graphene at 30~K and reported the appearance of a sharp $(\sqrt{3}\times\sqrt{3})R30^\circ$ diffraction pattern \cite{sugawara2011fabrication}, but did not observe this pattern after the deposition of Li on monolayer graphene. They suggested Li intercalates between the top two adjacent carbon layers and takes the same 
well-ordered superstructure as in bulk C$_6$Li. However, the band structure around the $\bar{K}$-point for both Li-deposited monolayer and bilayer graphene contained only a single $\pi$-band (see Figs.~2(e) and (f) in 
Ref.~[\onlinecite{sugawara2011fabrication}]). This contrasts significantly with earlier findings \cite{virojanadara2010epitaxial, virojanadara2010low, watcharinyanon2012studies} where at least two $\pi$-bands 
appear around the $\bar{K}$-point after Li deposition on monolayer graphene. 

We present here detailed ARPES data collected before and after Li deposition on 
monolayer graphene, both at room temperature and after heating to 
300$^\circ$C, in order to unambiguously determine the number of $\pi$-band 
branches present and their dispersions. Directly after 
deposition, we show clearly the presence of multiple $\pi$-bands around the 
$\bar{K}$-point. After heating the sample to 300$^\circ$C, these bands become 
considerably sharper and their dispersions do not resemble that of Bernal stacked bilayer 
graphene \cite{PhysRevLett.98.206802, RevModPhys.81.109}.  To understand these 
observations, tight-binding and density functional theory band structure calculations were 
performed for Li intercalated free-standing bilayer graphene, as well as for 
Li-intercalated graphene systems that explicitly take the SiC(0001) surface into 
account. 

Our combined experimental and theoretical results lead us to suggest the 
following: Directly after deposition, Li intercalates both underneath the buffer layer and between the two carbon layers. 
Heating the sample promotes the complete intercalation of the Li atoms to the 
interface, as well as the development of an ordered Li configuration between 
the two carbon layers. In the process, it becomes energetically more favorable 
for the carbon layers to become AA- rather than Bernal-stacked. We show 
that the $\pi$-bands visible around the $\bar{K}$-point closely resemble the 
band structure of a C$_6$LiC$_6$ system as calculated with density functional theory. The periodic perturbation of the graphene electronic structure by the ordered Li layer induces pseudo-`gaps' to open 
at the Dirac point of each of the $\pi$-cones. Tight-binding calculations show that the interlayer coupling is enhanced by the presence of the Li atoms beyond what is typical in clean bilayer graphene, and the impact of the enhanced Kekul\'{e}-textured skew-coupling terms on the band dispersion is verified by ARPES spectra. 

\section{Methodology}

\subsection{Experimental Details}

Monolayer graphene samples were grown in-situ on $n$-type wafers of 4H-SiC(0001) purchased from SiCrystal, which were specified to have a misorientation error within 0.05$^{\circ}$.
ARPES measurements were performed using Beamline I4 at the MAX IV Laboratory, which is equipped with a spherical grating monochromator and a PHOIBOS 100~mm CCD analyzer from SPECS. The wide angular dispersion mode was selected, providing an acceptant angle of $\pm$14$^{\circ}$.
Each ARPES spectrum was collected parallel to the 
$\bar{A}$-$\bar{K}$-$\bar{A}^\prime$ direction of the Brillouin zone of graphene, in steps of 0.25$^{\circ}$ along the $\bar{\Gamma}$-$\bar{K}$-$\bar{M}$ direction, using photon energies of $33$~eV and $70$~eV.
From this data, the $\pi$-band structure along certain directions in the Brillouin zone can be determined, as well as the angular distribution pattern, E$_i$(k$_x$,k$_y$), at selected initial state energies. A Ta foil was used as a reference to determine the Fermi level. 

Deposition of Li was performed during a five minute interval 
using a commercial alkali metal source (from SAES Getters) and with the sample at 
room temperature. Further details can be found in Refs.~[\onlinecite{virojanadara2010epitaxial, 
virojanadara2010low, watcharinyanon2012studies}]. Subsequent heating was carried out for four minutes at each selected temperature. The sample temperature was determined using optical pyrometers. ARPES measurements were taken after the sample had cooled to room temperature.
The Si 2p and Li 1s core level spectra were recorded before and after deposition and heating. They showed the same features and changes upon heating as reported earlier \cite{virojanadara2010epitaxial} and therefore no 
core level data is included below. 

\subsection{Computational Details}

Density functional theory calculations were performed using {\sc vasp}-5.3
\cite{Kresse1996, Kresse1999, PhysRevB.50.17953}. The Perdew-Burke-Ernzerhof 
(PBE) \cite{Perdew1996} parametrization of the generalized gradient 
approximation (GGA) was employed. The plane wave basis set was converged using 
an $800$~eV 
energy cutoff. Structural relaxations of the C$_6$LiC$_6$ cell were carried out 
using a $9\times9\times1$ $k$-point Monkhorst-Pack mesh \cite{PhysRevB.13.5188} 
to sample the Brillouin zone. A $24\times24\times1$ mesh was 
used to determine the total energies. The Tkatchenko-Scheffler method was used to describe van der Waals interactions \cite{PhysRevLett.102.073005}. The 
free-standing bilayer structures were optimized until all residual forces were 
less than \SI{0.001}{\electronvolt \per \angstrom}

The \sic\ substrate was modelled using an asymmetric slab consisting of 6 
bilayers of \sic, arranged in the ABCACB stacking associated with the 6H 
polytype. A bulk termination was assumed. The GGA calculated lattice constant of 
bulk SiC is \SI{3.09}{\angstrom}, in good agreement with the experimental value 
of \SI{3.08}{\angstrom}. A vacuum layer of at least \SI{15}{\angstrom} was 
included in the direction normal to the surface to ensure no spurious 
interactions between repeating slabs and the dipole correction was applied where 
appropriate. The dangling C bonds on the SiC($000\bar{1}$) surface were passivated 
with H atoms. The positions of the top two bilayers of \sic, as well as the 
H-terminating atoms, the Li atoms, and all carbon layers, were optimized until 
all residual forces were less than \SI{0.01}{\electronvolt \per \angstrom}. The 
remaining atoms were held fixed at their bulk positions. 
The ($6\sqrt{3}\times6\sqrt{3})R30^\circ$ surface reconstruction 
\cite{owman1996sic, chen2005atomic, PhysRevB.45.1327, hass2006highly} of 
the buffer layer on SiC(0001) was modelled using the simplified 
$(\sqrt{3}\times\sqrt{3})R30^\circ$ cell which corresponds to a $2 \times 2$ 
graphene cell. Such an approximation was shown to be adequate to correctly 
describe the interaction between the \sic\ surface and the carbon layers 
\cite{PhysRevB.77.155303}.

\section{Results}

\subsection{Experimental ARPES Spectra}
\begin{figure}[ht]
\begin{centering}
\includegraphics[width=\linewidth]{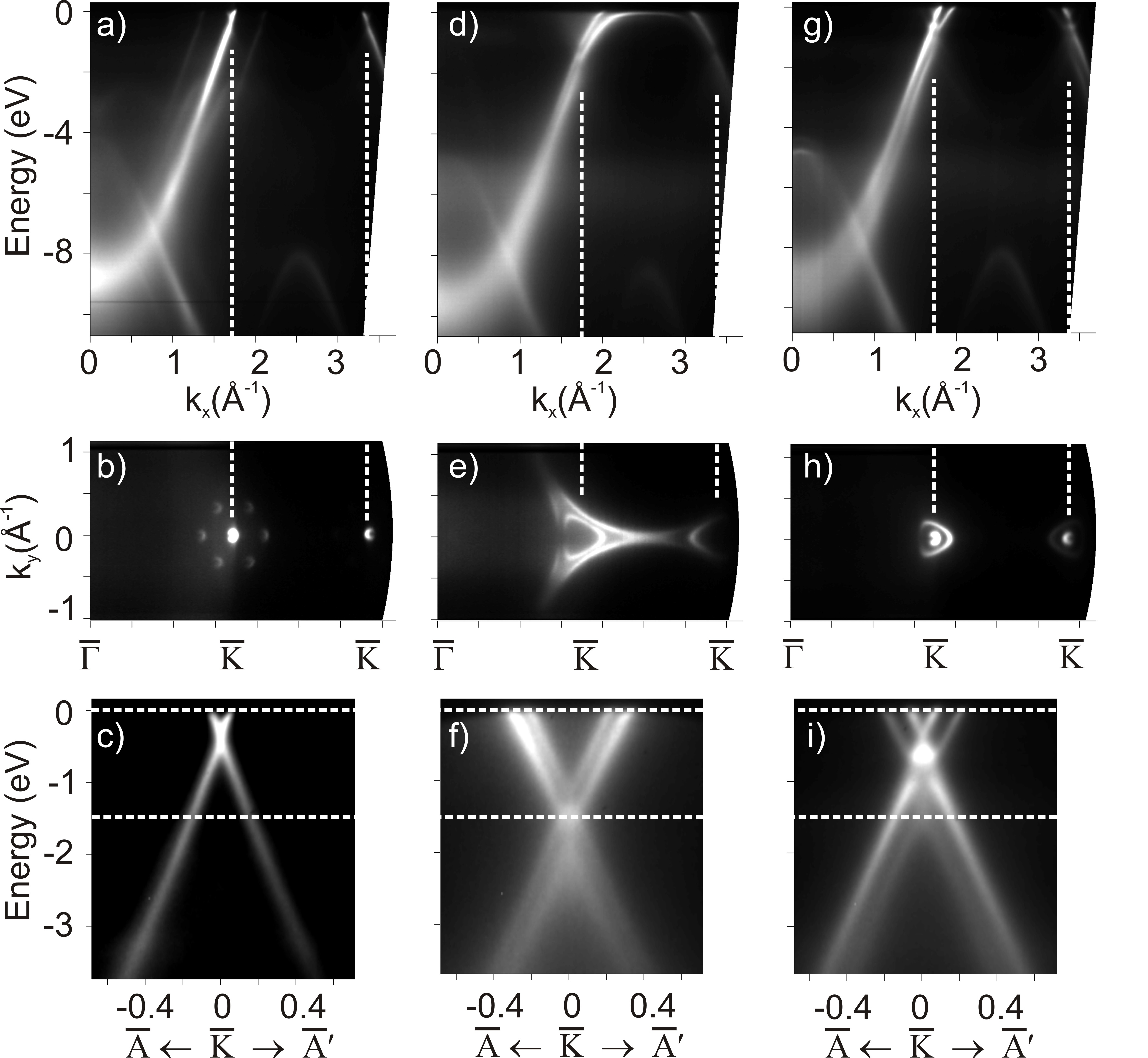}
\caption{\label{fig:exp} (Color online) Band dispersion recorded from monolayer graphene 
samples before (a -- c) and after (d -- f) Li deposition, and after heating to 
300$^\circ$C (g -- i). The dashed lines are guides to the eye for 
illustrating the location of the first and second $\bar{K}$ points or an initial state energy of $0$~eV and $-1.4$~eV. The Fermi energy is located at 0~eV. $\bar{A}$ denotes the direction perpendicular to the  $\Gamma$-$\bar{K}$-$\bar{M}$ direction in the 2D Brillouin zone of graphene.}
\end{centering}
\end{figure}

The band structure obtained from the as-grown monolayer graphene sample, along the 
$\bar{\Gamma}$-$\bar{K}$-$\bar{M}$ direction of the Brillouin zone of graphene, 
is displayed in Fig.~\ref{fig:exp}(a). 
The contribution from the $\pi$-band dominates, with branches visible at both the first and second $\bar{K}$ point. The photoelectron angular distribution pattern, E$_i$($k_x,k_y$), obtained at the 
Fermi energy is shown in Fig.~\ref{fig:exp}(b). The image has been overexposed to show the presence of six replica $\pi$-cones surrounding the Dirac cone at the first $\bar{K}$ 
point, as well as the presence of the Dirac cone at the second $\bar{K}$ 
point. The Dirac cone at the second $\bar{K}$ point appears, as it should, as a 
mirror image of the one at the first $\bar{K}$ point. The lower intensity obtained around 
the second $\bar{K}$-point is attributed to photoelectron matrix effects 
\cite{PhysRevB.83.121408, PhysRevB.77.195403}. 
The $\pi$-band structure obtained along the $\bar{A}$-$\bar{K}$-$\bar{A}^\prime$ direction around the first 
$\bar{K}$ point is shown in Fig.~\ref{fig:exp}(c). A single, linearly dispersing 
$\pi$-band is visible with the Dirac point located approximately $0.45$~eV below the 
Fermi level. Taken together, the ARPES spectra in Figs.~\ref{fig:exp}(a), (b) 
and (c) demonstrate that high quality, monolayer graphene is present on the 
SiC(0001) surface \cite{PhysRevLett.98.206802}.

Depositing Li on the sample at room temperature induces significant changes in 
the band structure, as illustrated in Figs.~\ref{fig:exp}(d), (e) and (f). 
Fig.~\ref{fig:exp}(d) shows that the entire $\pi$-band structure, here displayed 
in the $\bar{\Gamma}$-$\bar{K}$-$\bar{M}$ direction, is rigidly shifted downwards by about $1$~eV; the position 
of the $\pi$-band at the $\bar{\Gamma}$-point is now located at $-10$~eV, whereas for 
the clean sample it is located at approximately $-9$~eV [c.f. 
Fig.~\ref{fig:exp}(a)]. A second $\pi$-band is now visible close to the Fermi 
energy and the dispersion of these bands is no longer linear. The lower branch 
is instead quite flat and located below the Fermi level. This can be seen 
clearly in the Fermi surface [Fig.~\ref{fig:exp}(e)] where the 
$\pi$-bands are shown to be occupied all the way between the first and second 
$\bar{K}$-points. Again, the Fermi surface near the second 
$\bar{K}$-point is a mirror image of that near the first. 
The $\pi$-bands in the $\bar{A}$-$\bar{K}$-$\bar{A}^\prime$ direction are shown in Fig~\ref{fig:exp}(f). The  Dirac point has now shifted down in energy to approximately $1.4$~eV below the Fermi level. This is due to the strong $n$-type electron doping provided by the Li atoms to the graphene layers. 

Heating the sample to 300$^\circ$C induces further changes in the 
band structure, as illustrated in Figs.~\ref{fig:exp}(g), (h) and (i). The 
entire band structure, displayed in Fig.~\ref{fig:exp}(g), is now rigidly 
shifted by about $0.8$~eV back towards the Fermi level. This can be seen by comparing the $\pi$-bands at the 
$\bar{\Gamma}$-point in (g) and (d). Two $\pi$-band branches are now clearly resolved in an energy window between the Fermi level and $-6$~eV. The Fermi surface in Fig.~\ref{fig:exp}(h) shows that neither 
of these $\pi$-band branches is continuously located below the 
Fermi level between the first and second $\bar{K}$-point. 
Again, the Fermi surfaces at the first and second $\bar{K}$-points are mirror images of each other. 

Comparing Figs.~\ref{fig:exp}(i) and (f), one can see that the $\pi$-bands 
around the $\bar{K}$-point are considerably sharper, and their dispersions 
distinctly different, after heating than directly after Li deposition at room 
temperature. The Dirac point has moved to approximately $0.6$~eV below the Fermi level with two distinct $\pi$-bands visible above and below it. The bands show neither the parabolic dispersion of Bernal stacked bilayer graphene nor the linear dispersion predicted for AA-stacking \cite{PhysRevLett.98.206802, RevModPhys.81.109}.

\begin{figure}[ht]
\includegraphics[width=0.85\linewidth]{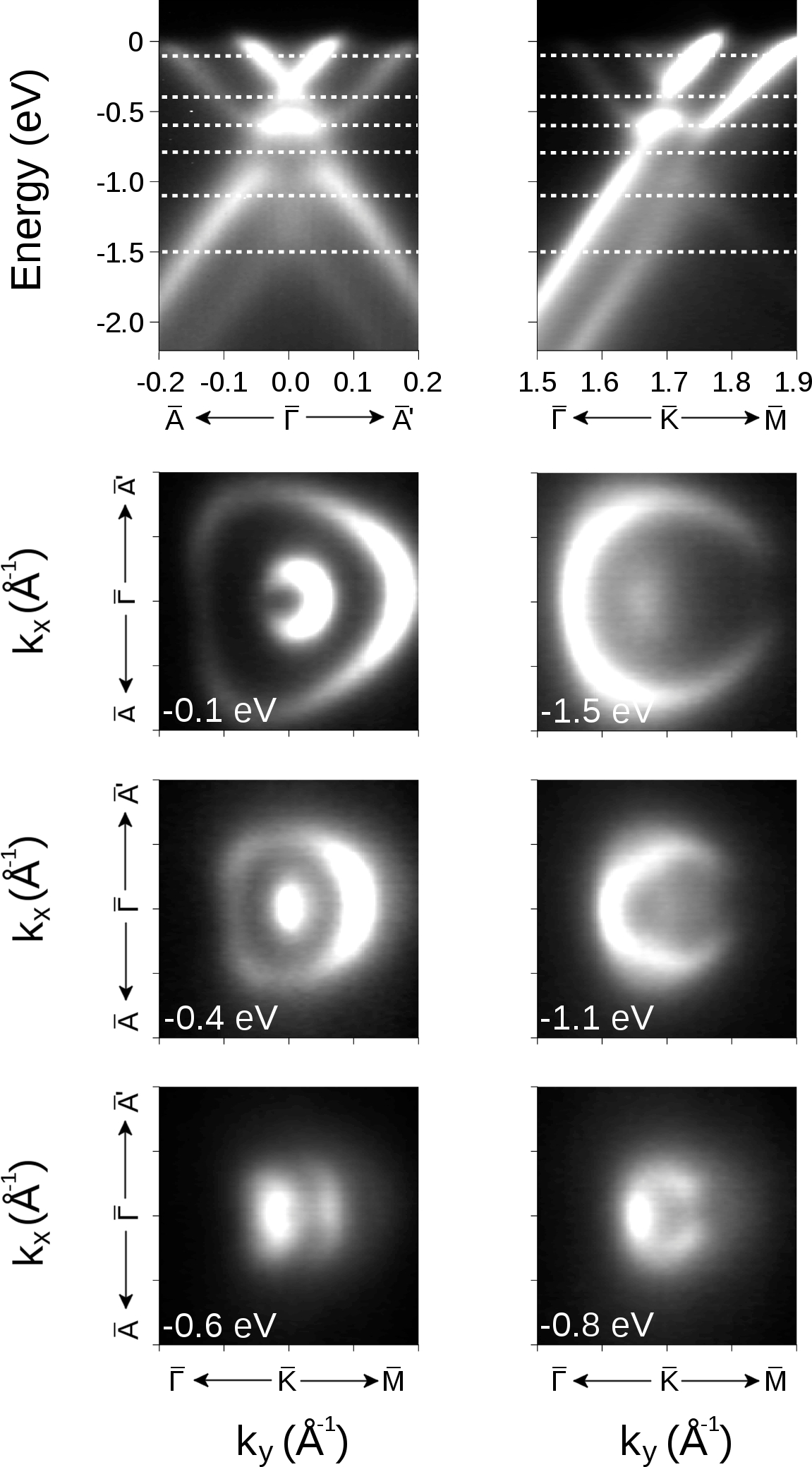}
\caption{\label{fig:angular_dispersions} (Color online) Band dispersion recorded in the vicinity of the $\bar{K}$-point (top two panels) and angular distributions extracted at fixed energies close to the Dirac point. The six energies are shown as white dashed lines through the band dispersions in the top panels. The Fermi energy is located at 0~eV. An incident photon energy of 33~eV was used. }
\end{figure}

Photoemission angular distribution patterns extracted at several fixed energies above and below the Dirac point are shown in Fig.~\ref{fig:angular_dispersions}. 
At energies far from the Dirac point, namely at $-0.1$~eV and $-1.5$~eV from the Fermi energy, two ring-like patterns corresponding to the two $\pi$-bands are present. A strong triangular deformation, or trigonal warping, of the outer Dirac cone is clearly visible at these energies. 
At energies closer to the Dirac point, for example at $-0.4$~eV and $-1.1$~eV, the trigonal warping is considerably reduced and the inner ring has been reduced to a point. Finally, the angular distribution at energies of $-0.6$~eV and $-0.8$~eV are not comprised of ring-like segments but of two and three points of varying intensity, respectively. In all cases, the intensity anisotropy of the angular maps does not correspond to that predicted for Bernal-stacked bilayer graphene \cite{PhysRevB.77.195403, hwang2011direct}.

\subsection{DFT Band Structures}
The ARPES spectra presented in the previous section shows clearly the presence of two $\pi$-bands after Li deposition, which implies that Li has intercalated to the interface and decoupled the buffer layer from the substrate. This has previously been corroborated by both experimental core-level spectra and DFT calculations and verified in the current work (see Appendix). We will now address the relationship between the presence of Li between the two carbon layers and its effect on the dispersion by calculating the band structure of free-standing bilayer graphene with and without intercalated Li.

\begin{figure}[ht]
\begin{centering}
\includegraphics[width=0.8\linewidth]{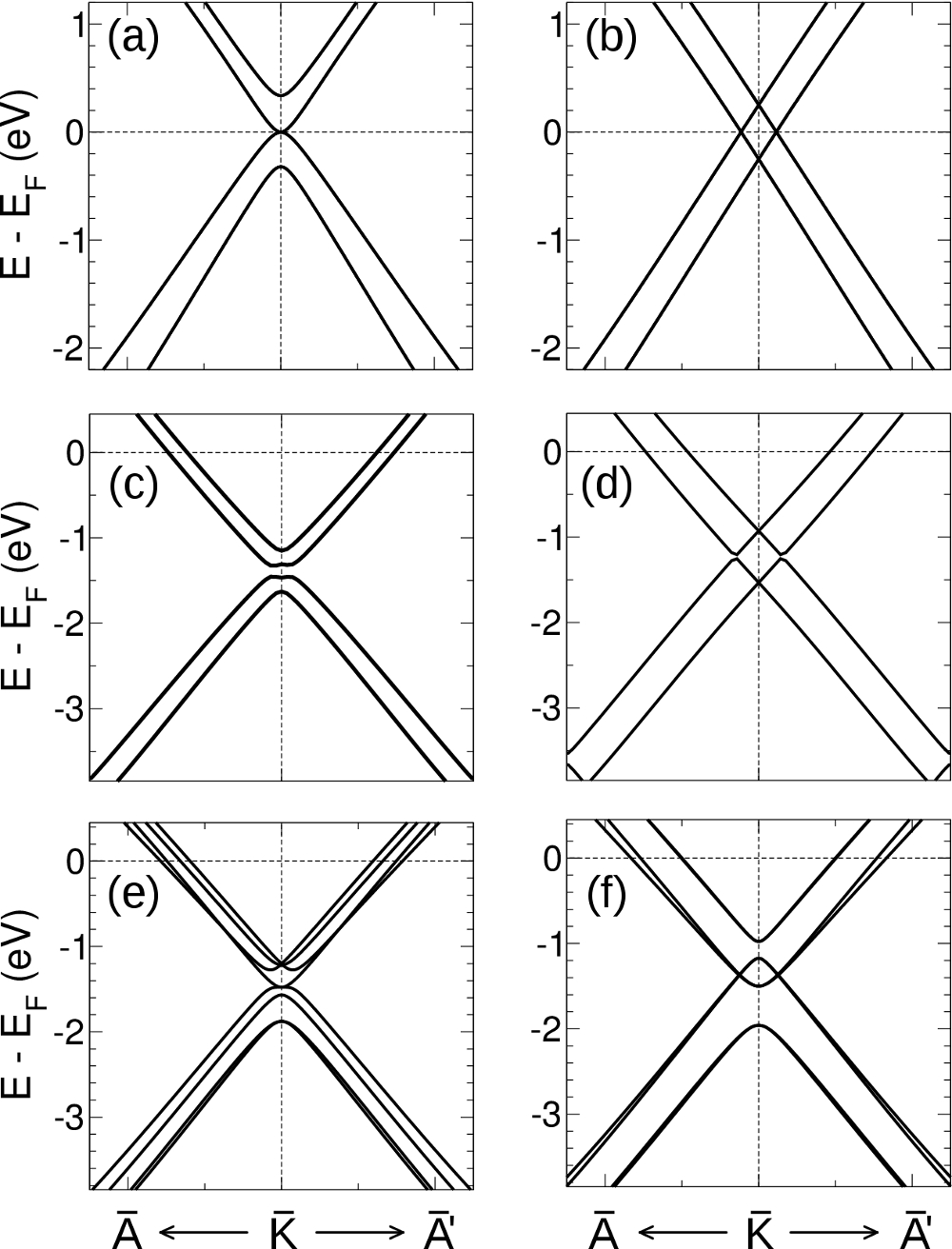}
\caption{\label{fig:theory} Calculated band structure of (a) AB-stacked clean 
bilayer graphene, (b) AA-stacked clean bilayer graphene (c) AB-stacked C$_8$LiC$_8$, (d) AA-stacked C$_8$LiC$_8$, (e) AB-stacked C$_6$LiC$_6$ and (f) AA-stacked C$_6$LiC$_6$. The high-symmetry labels are those of the graphene unit cell.}
\end{centering}
\end{figure}

The first possibility is that Li does not intercalate between the two carbon layers. These two layers then remain Bernal stacked and the resulting band structure will have a symmetric, parabolic band dispersion around the $\bar{K}$-point, as shown in Fig.~\ref{fig:theory}(a). 
The band structure of clean AA-stacked bilayer graphene, shown in Fig.~\ref{fig:theory}(b), is comprised of two 
Dirac cones with linear dispersion, separated in energy by $0.5$~eV. We find an interlayer distance of 
3.52~\AA\ for clean bilayer AA-stacked graphene, close to the experimentally found value for 
AA-stacked graphite \cite{lee2008growth}. 

A second possibility is that Li intercalates between the two carbon layers, which remain AB-stacked, such that a Li atom sits on top of the centre of a hexagon for one graphene layer and directly underneath a carbon atom of the second carbon layer. The band structure associated with such a scenario is shown in Fig.~\ref{fig:theory}(c) for a C$_8$LiC$_8$ concentration.
The structural asymmetry is evident in the resulting `Mexican hat' band dispersion. A gap opens at the $\bar{K}$-point as a result of the asymmetric doping of the two carbon layers. For this particular Li concentration a gap of $0.13$~eV is opened. 
The third possibility is that the registry between the two graphene layers switches from AB- to AA-stacking. 
A sufficiently high concentration of intercalated Li atoms is already known to shift the stacking pattern of bilayer graphene and Li-graphite intercalation compounds (Li-GICs) to an AA-stacking of the carbon sheets \cite{imai2007energetic,zhou2012first}. The resulting band structure is shown in Fig.~\ref{fig:theory}(d). It has the linear dispersion typical of AA-stacked bilayer graphene. The effect of the intercalated Li can be seen in the increased energy separation in the two Dirac cones, from 0.5~eV to 0.6~eV. 

The highest concentration of Li found in GICs at ambient conditions is LiC$_6$. The $(\sqrt{3}\times\sqrt{3})R30^\circ$ diffraction pattern observed after Li deposition 
at room temperature \cite{virojanadara2010epitaxial} would suggest that this Li concentration is also the most favourable in bilayer graphene. 
If we consider AB-stacked graphene with a C$_6$LiC$_6$ concentration, we find three $\pi$-bands are visible, due to the asymmetric doping of the two graphene layers.
Finally, we consider AA-stacked graphene at the same concentration.
The energy dispersion, shown in Fig.~\ref{fig:theory}(f), is comprised of two Dirac cones shifted relative to one another by the interlayer coupling and with a pseudo-gap opened in both at the $\bar{K}$-point. 
We find that AA-stacked C$_6$LiC$_6$ is 36~meV/C lower in energy than the equivalent AB-stacked configuration. As a comparison, Bernal stacked clean bilayer graphene is 6~meV/C lower in energy than AA-stacked bilayer graphene. 

\section{Discussion}

\begin{figure}[ht!]
\begin{centering}
\includegraphics[width=0.9\linewidth]{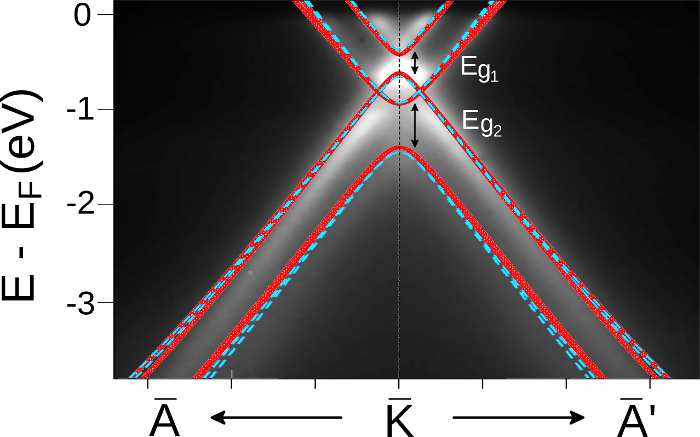}
\caption{\label{fig:compare} (Color online) Band structure of free-standing C$_6$LiC$_6$ as calculated with DFT (red solid lines) and using a tight-binding model (blue dashed lines), compared to the experimental ARPES spectra [c.f. Fig.~\ref{fig:exp}(i)]. The calculated band structures have been shifted rigidly in energy to match the top $\pi^*$-band.}
\end{centering}
\end{figure}

Comparing the band structures presented in Fig.~\ref{fig:theory} with the experimental ARPES spectra in Fig.~\ref{fig:exp}(i), we find that agreement is only achieved if we assume a C$_6$LiC$_6$ concentration of Li atoms between AA-stacked bilayer graphene, as shown in Fig.~\ref{fig:compare}. 
The development of severe cracks and wrinkles after Li deposition \cite{virojanadara2010low} further supports this postulation. We suggest that this is due to the strain induced in the system when the top carbon layer attempts to shift to an AA-stacking. These cracks do not appear when Li is deposited on the buffer layer only.

\subsection{Tight-binding Model}

\begin{figure}[ht!]
\begin{centering}
\includegraphics[width=0.95\linewidth]{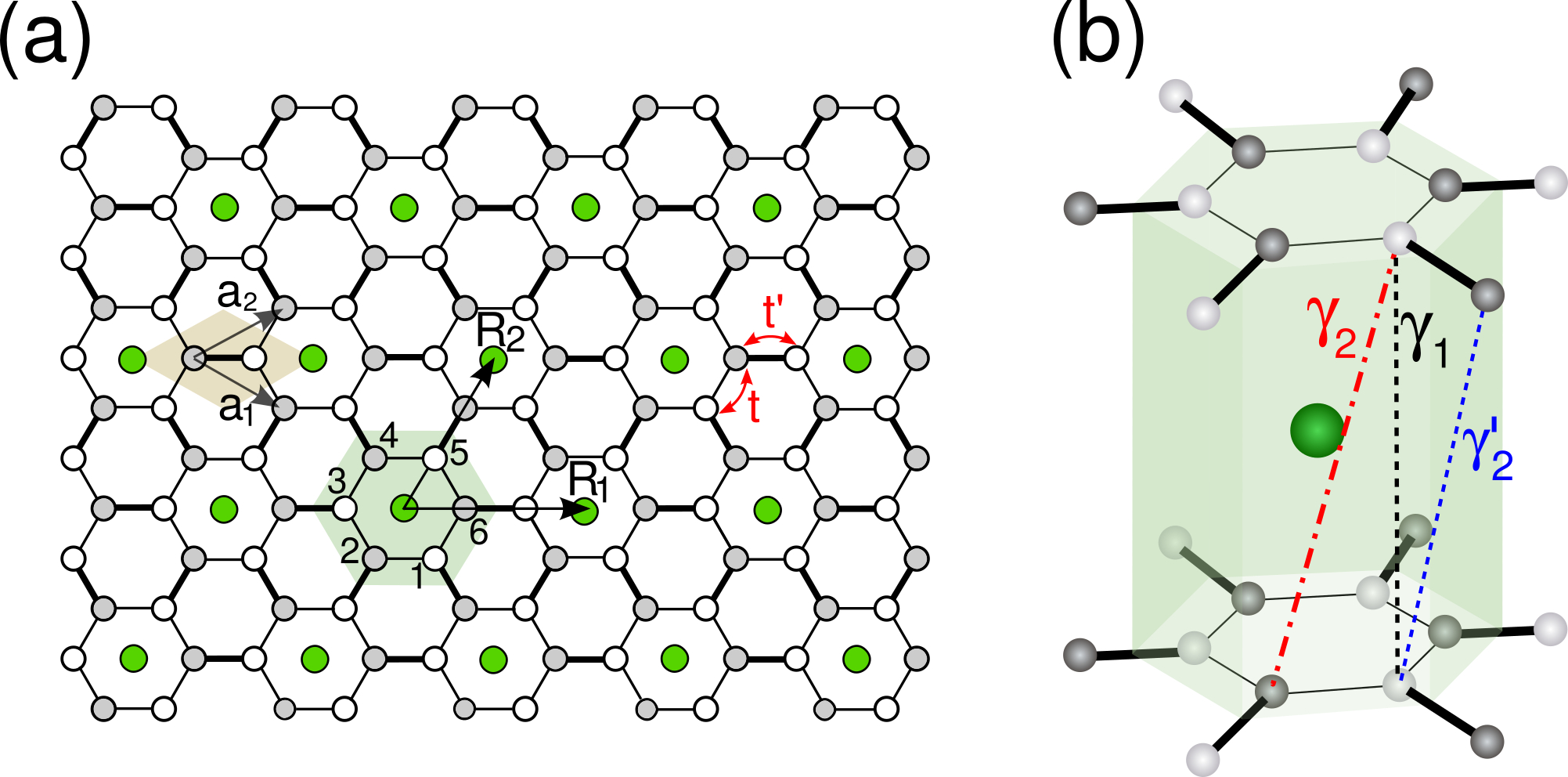}
\caption{\label{fig:theory_symmetry} (Color online) Structure of C$_6$LiC$_6$. (a) Top view 
showing the graphene lattice vectors $\vec{a}_1$ and $\vec{a}_2$ and the lattice 
vectors of C$_6$LiC$_6$, $\vec{R}_1$ and $\vec{R}_2$. $\vec{R}_1$ 
and $\vec{R}_2$ are rotated by 30$^\circ$ with respect to $\vec{a}_1$ and 
$\vec{a}_2$ and are $\sqrt{3}$ longer. The two carbon sublattices are denoted with gray and white circles and Li atoms are shown as green circles. The two types of bonds associated with the Kekul\'{e} textured graphene are shown as thin and thick solid lines. The two different intralayer hopping parameters between carbon atoms associated with these two bonds are denoted $t$ and $t^\prime$. The 6 atoms belonging to the C$_6$LiC$_6$ unit cell of one of the graphene layers are labeled 1-6, and are used to generate the Hamiltonian of the system. (b) Side view showing the interlayer coupling parameters, $\gamma_1$ between carbon atoms directly on top of one another, and $\gamma_2$ and $\gamma_2^\prime$ describing the textured skew coupling terms.}
\end{centering}
\end{figure}

We will now discuss the unusual shape of the electronic band dispersion close to the Dirac point, and in particular the opening of the two pseudo-gaps of differing magnitudes at the $\bar{K}$-point, using a simple tight-binding model. 
The Li atoms, which intercalate in an ordered fashion, cause a periodic modification of the graphene electronic structure. The effect can be described by a periodic potential $U(\vec{r}) = U(\vec{r} + l_1\vec{R}_1 + 
l_2\vec{R}_2)$ where $\vec{R}_1 = n_1\vec{a}_1 + m_1\vec{a}_2$, $\vec{R}_2 = n_2\vec{a}_1 + m_2\vec{a}_2$, and $\vec{a}_1$ and $\vec{a}_2$ are the graphene lattice vectors \cite{dvorak2013bandgap}. $l$, $n$ and $m$ are integers. In our case, $\vec{R}_1$ and $\vec{R}_2$ describe 
the lattice vectors of C$_6$LiC$_6$ as shown in Fig.~\ref{fig:theory_symmetry}(a). Here, $\vec{a}_1 = (\frac{\sqrt{3}}{2}, -\frac{1}{2}) a_0$, $\vec{a}_2 = (\frac{\sqrt{3}}{2}, \frac{1}{2}) a_0$, $\vec{R}_1 = (\sqrt{3}, 0) a_0$ and $\vec{R}_2 = (\frac{\sqrt{3}}{2}, \frac{3}{2}) a_0$, where $a_0$ is the graphene lattice constant. 
A band gap will open at the $\bar{K}$-point when the periodic potential in reciprocal 
space, $U(K)$, is not equal to zero. This condition will be satisfied when $(n_1 - m_1)\bmod~3 
= 0$ and $(n_2 - m_2)\bmod~3 = 0$. This rule is reminiscent of that describing 
the gap chirality in carbon nanotubes \cite{reich2008carbon} and the effect has already
been harnessed to open bandgaps in monolayer and bilayer graphene through the periodic 
pattering of antidot lattices \cite{petersen2010clar, 
PhysRevB.91.115424}. The triangular symmetry associated with a $(\sqrt{3}\times\sqrt{3})R30^\circ$  Li adsorption pattern has already been shown to satisfy the conditions necessary to open a gap in monolayer graphene at the Dirac point \cite{PhysRevLett.106.187002, PhysRevB.79.045417}.

We will now extend the model of Ref.~[\onlinecite{PhysRevB.79.045417}] from Li decorated monolayer graphene to AA-stacked bilayer C$_6$LiC$_6$ graphene using a tight-binding model that takes the effect of the intercalated Li into account via a Kekul\'{e} texturing of the nearest neighbour hopping parameters.
The 12$\times$12 matrix that describes the Hamiltonian of such as system has the following form:
$$ H=\begin{pmatrix}
 H_{\mathrm{T}} & H_{\mathrm{TB}}\\
H_{\mathrm{BT}} & H_{\mathrm{B}}
\end{pmatrix}
$$
where $H_{\mathrm{T(B)}}$ is the 6$\times$6 matrix describing intralayer hopping 
in the top (bottom) graphene layer and $H_{\mathrm{TB}}$ is the 6$\times$6 
matrix describing interlayer coupling. 
The intralayer Hamiltonian, $H_{\mathrm{T}} = H_{\mathrm{B}}$, is given by:
$$\Scale[0.92]{
\begin{pmatrix}
 \epsilon & t & 0 & t^\prime e^{i \vec{k}.\vec{\tau_1}} & 0 & t\\
 t & \epsilon & t & 0 & t^\prime e^{-i \vec{k}.\vec{\tau_2}} & 0\\
 0 & t & \epsilon & t & 0 & t^\prime e^{i \vec{k}.\vec{\tau_3}}\\
 t^\prime e^{-i \vec{k}.\vec{\tau_1}} & 0 & t & \epsilon & t & 0\\
 0 & t^\prime e^{i \vec{k}.\vec{\tau_2}} & 0 & t & \epsilon & t\\
 t & 0 & t^\prime e^{-i \vec{k}.\vec{\tau_3}} & 0 & t & \epsilon
\end{pmatrix}}
$$
where 
$t$ and $t^\prime$ are the textured nearest neighbour carbon hoppings (see Fig.~\ref{fig:theory_symmetry}(a)), $\epsilon$ is the on-site energy and
$ \vec{\tau_{1,2}} =  a_0\left(\frac{\sqrt{3}}{2}, \mp\frac{3}{2}\right) 
\hspace{4pt} \mathrm{and} \hspace{6pt} \vec{\tau_3} = a_0 \left(-\sqrt{3}, 
0\right)$.
The interlayer coupling matrix, $H_{\mathrm{TB}} = H_{\mathrm{BT}}$, is given by:
$$\Scale[0.92]{\begin{pmatrix}
 \gamma_1 & \gamma_2 & 0 & \gamma_2^\prime e^{i \vec{k}.\vec{\tau_1}} & 0 & 
\gamma_2\\
 \gamma_2 & \gamma_1 & \gamma_2 & 0 & \gamma_2^\prime e^{-i 
\vec{k}.\vec{\tau_2}} & 0\\
 0 & \gamma_2 & \gamma_1 & \gamma_2 & 0 & \gamma_2^\prime e^{i 
\vec{k}.\vec{\tau_3}}\\
 \gamma_2^\prime e^{-i \vec{k}.\vec{\tau_1}} & 0 & \gamma_2 & \gamma_1 & 
\gamma_2 & 0\\
 0 & \gamma_2^\prime e^{i \vec{k}.\vec{\tau_2}} & 0 & \gamma_2 & \gamma_1 & 
\gamma_2\\
 \gamma_2 & 0 & \gamma_2^\prime e^{-i \vec{k}.\vec{\tau_3}} & 0 & \gamma_2 & 
\gamma_1 
\end{pmatrix}}
$$
where 
$\gamma_1$ is the vertical interlayer coupling between two carbon atoms directly on top of one another and $\gamma_2$ and $\gamma_2^\prime$ are the textured skew hopping terms. These are sketched in Fig.~\ref{fig:theory_symmetry}(b).
The resulting band structure is shown by blue dashed lines in Fig.~\ref{fig:compare}.
We find that, as a result of the intercalation symmetry, a pseudo-gap opens at the Dirac point of each cone. These two cones are then shifted in energy relative to one another with an energy equal to the vertical interlayer couping, $\gamma_1 = 0.34$~eV. 
The in-plane hopping parameters, $t=2.79$~eV and $t^\prime=2.59$~eV, agree with those found in in Ref.~[\onlinecite{PhysRevB.79.045417}] to describe Li-adsorbed monolayer graphene. The values of $\gamma_1=-0.34$~eV and $\gamma_2^\prime = 0.04$~eV are characteristic of AA-stacked bilayer graphene and graphite \cite{charlier1992first}. 

In contrast to clean AA-stacked graphene or graphite, however, the effect of the skew coupling terms cannot be neglected. The effect of their textured nature on the band structure can be seen in the breaking of the equivalence of the two pseudo-gaps. If $\gamma_2$ and $\gamma_2^\prime$ were equal, the pseudo-gap opened in each cone would also be equal. Instead we have that one gap, E$_{\mathrm{g_{1}}}$, is $0.20$~eV wide, while the other, E$_{\mathrm{g_{2}}}$, is 2.3 times larger, at $0.46$~eV. This ratio is approximately equal to that between $\gamma_2$ and $\gamma_2^\prime$ which are 0.10~eV and 0.04~eV, respectively.

To conclude, we have shown that Li intercalation serves to switch the stacking in a graphene bilayer from AB to AA. The highly ordered Li layer generates a periodic perturbation of the graphene electronic structure such that a pseudo-gap opens in each cone at the Dirac point. The increase in magnitude of the skew coupling terms is significant and its effects are visible in the experimental ARPES spectra. 

\subsection{Reduction in Doping after Heating}

\begin{figure}[ht!]
\begin{centering}
\includegraphics[width=0.95\linewidth]{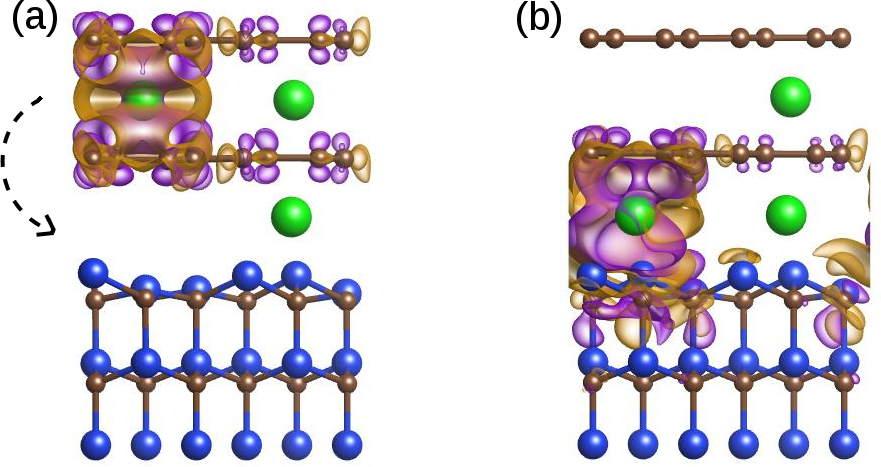}
\caption{\label{fig:top_bottom} (Color online) Side view of a 2x1 unit cell of SiC(0001) that includes two layers of graphene (16 C atoms in each) and three Li atoms. Isosurfaces of charge density difference show how charge is transferred from a Li atom inserted (a) between the two carbon layers and (b) at the interface between SiC and the bottom carbon layer. A purple (orange) isosurface refers to a gain (loss) in charge density. The isosurface value is 0.0005~e/a.u.$^3$.}
\end{centering}
\end{figure}

Finally, we discuss the observation that the graphene doping level decreases significantly, from $-1.4$~eV to $-0.6$~eV, after heating to 300$^\circ$C [c.f. Fig.~\ref{fig:exp}(f) and (i)]. 
Core-level spectra shows that this dramatic reduction in doping occurs despite evidence that almost half of the Li atoms have not deintercalated or desorbed from the sample. 
The Li 1s core level spectrum shows that the component assigned to surface-adsorbed Li essentially disappears while the components assigned to intercalated Li remain, in agreement with earlier observations \cite{virojanadara2010epitaxial, virojanadara2010low, watcharinyanon2012studies, PhysRevB.91.245411}. We therefore suggest that some Li atoms which were initially adsorbed on the surface, or between the two graphene layers, intercalate to the interface between SiC and graphene after heating. In doing so, part of the charge which was previously transferred to the carbon atoms is now transferred to the SiC substrate, thus reducing the graphene doping level. 
If we consider a $2\times1$ SiC(0001) unit cell, that includes two graphene layers and three Li atoms as sketched in Fig.~\ref{fig:top_bottom}, we find that it is energetically more favourable, by 0.54~eV, for one of the Li atoms to move from between the two graphene layers to the interface between the substrate and the bottom graphene layer. 

To visualize the charge transfer, we show in Fig.~\ref{fig:top_bottom}(a) the charge density difference (CDD) that occurs when a Li atom is inserted between the top two carbon layers. The CDD is defined as $\Delta\rho = \rho_{\mathrm{final}} - \rho_{\mathrm{inital}} - \rho_{\mathrm{Li}}$ , where $\rho_{\mathrm{final}}$, $\rho_{\mathrm{inital}}$ and $\rho_{\mathrm{Li}}$ are the charge densities of the final system, the initial system that includes only two Li atoms, and an isolated Li atom, respectively. 
The charge is transferred equally between the two carbon layers. In contrast, Fig.~\ref{fig:top_bottom}(b) shows how the charge is distributed from the Li atom when it is placed at the interface between SiC and the buffer layer. In this case, the majority of the charge is transferred to the substrate, with only a small amount transferred to the decoupled buffer layer. This process can explain the shifting of the Dirac cones towards the Fermi level after heating as observed in the ARPES spectra.

\section{Conclusion}

We have carried out detailed ARPES (Angle Resolved Photoelectron Spectroscopy) 
studies of the band structure of monolayer graphene samples, before and after Li deposition at room 
temperature as well as after subsequent heating to 300$^\circ$C. Li intercalation is shown to have a large impact on the dispersion of the bands at energies close to the Dirac point, as well as donating charge to the graphene layers. Directly after Li deposition we observe the appearance of a second $\pi$-band. Upon heating to 300$^\circ$C the electronic band structure changes significantly: The $\pi$-bands become significantly sharper and have a distinctly different dispersion to that of Bernal stacked bilayer graphene. The position of the Dirac point is also shifted closer to the Fermi energy to approximately $-0.6$~eV. To understand these observations, we performed density functional theory band structure calculations for Li intercalation in free-standing bilayer graphene for different stacking of the layers, as well as for Li-intercalated graphene on the SiC(0001) surface. Our combined experimental and theoretical results show that, after Li deposition and subsequent heating, Li both intercalates underneath the 
buffer layer -- decoupling it from the substrate -- and between the two carbon layers. In the process, it becomes energetically more favorable for the carbon layers to become AA-stacked rather than Bernal stacked. We show that the $\pi$-bands around the $\bar{K}$-point closely resemble the calculated band structure of a C$_6$LiC$_6$ system, where 
the intercalated Li atoms introduce a periodic perturbation to the graphene electronic structure that opens pseudo-gaps in the $\pi$-bands at the Dirac point.

\begin{acknowledgments}
The authors wish to acknowledge the Swedish Research Council (VR) grants No. 
621-2011-4252 and 621-2011-4426, the Swedish Foundation for Strategic Research 
(SSF) program  SRL Grant No. 10-0026, the European Union Seventh Framework 
Programme under grant agreement no. 604391 Graphene Flagship.
R.A. acknowledges financial support from VR grant No. 621-2011-4249 and the 
Linnaeus Environment at Link\"oping on Nanoscale Functional Materials (LiLi-NFM) 
funded by VR.
I.A.A. acknowledges the support from the Grant of Ministry of Education and 
Science of the Russian Federation (Grant No. 14.Y26.31.0005) and Tomsk State 
University Academic D. I. Mendeleev Fund Program (project No. 8.1.18.2015).
All calculations were performed using the supercomputer resources of the Swedish 
National Infrastructure for Computing (SNIC) National Supercomputing Center 
(NSC) and the PDC Centre for High Performance Computing (PDC-HPC).

\end{acknowledgments}

\appendix*

\section{Li-intercalated graphene on SiC(0001)}
 
To determine the effect of doping due to the substrate, and also to verify that Li intercalation is sufficient to decouple the carbon buffer layer from the Si surface, we now take the SiC(0001) surface into account explicitly. Due to computational constraints we could not model a C$_6$LiC$_6$ bilayer graphene on top of SiC and therefore cannot reproduce the pseudo-gaps that open as a result of this periodic potential. 
Instead we model a C$_8$LiC$_8$ concentration.  In agreement with the results of Refs.~[\onlinecite{PhysRevB.91.245411}] and [\onlinecite{li2011lithium}] we find that Li intercalation at the interface is sufficient to decouple the carbon buffer layer from the substrate. The Li atom is located at the T4 site on the SiC(0001) surface, breaking the Si -- C bonds. The decoupled graphene sheet is then free to move laterally so that the centre of the carbon hexagon is directly over the Li atom beneath it. 

\begin{figure}[h!]
\begin{centering}
\vspace{\baselineskip}
\includegraphics[width=\linewidth]{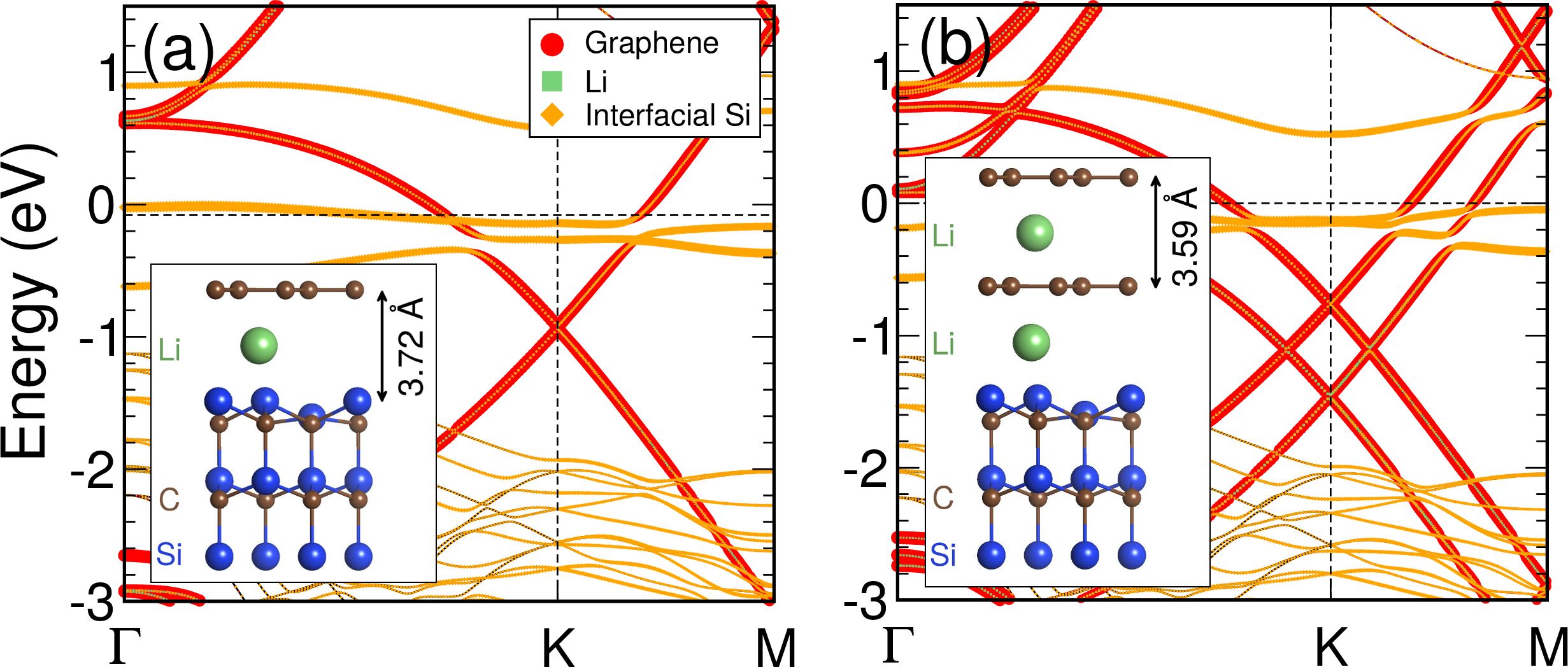}
\caption{\label{fig:theory_sic} (Color online) Electronic band structure of (a) the Li 
intercalated zero layer graphene on SiC(0001) system and (b) Li intercalated 
monolayer graphene on SiC(0001). The Li concentration corresponds to LiC$_8$ and 
LiC$_8$LiC$_8$, respectively. The insets show the relaxed structure of both 
configurations. The Fermi energy is located at 0~eV.}
\end{centering}
\end{figure} 

The band structure for the configuration that includes only zero layer graphene is 
shown in Fig.~\ref{fig:theory_sic}(a). A single $\pi$-band, associated with the now decoupled carbon layer, is visible. The Dirac point is located $0.93$~eV below the Fermi energy due to the strong doping from the Li atom.  
Fig.~\ref{fig:theory_sic}(b) then shows the band structure of the configuration 
that includes a second carbon layer. Li atoms are now present both underneath the buffer layer and between the two carbon layers. As in the free-standing case, the AA-stacked configuration is energetically preferred. Two Dirac cones are now visible, separated in energy by 0.7~eV. This would suggest an interlayer coupling, $\gamma_1$, of approximately 0.35~eV, similar to what we found for the free-standing intercalated bilayer. As a result of the doping due to these Li atoms, the Dirac 
points of the two cones are located $1.5$~eV and $0.8$~eV below the Fermi level, 
respectively.

%

\end{document}